\documentclass[12pt]{article}

\usepackage{amsmath,amssymb,graphicx} %use drftcite in draftmode
\usepackage{epsf}
\usepackage{pstricks}
\usepackage{cite}

\newcommand{\eq}[1]{~{\rm (\ref{eq:#1})}}

\newcommand{\beq}{\begin{eqnarray}}% can be used as {equation} or  {eqnarray}
\newcommand{\eeq}{\end{eqnarray}}

\newcommand{\centeron}[2]{{\setbox0=\hbox{#1}\setbox1=\hbox{#2}\ifdim                
\wd1>\wd0\kern.5\wd1\kern-.5\wd0\fi
\copy0

\kern-.5\wd0\kern-.5\wd1\copy1\ifdim\wd0>\wd1
                                       \kern.5\wd0\kern-.5\wd1\fi}}
\newcommand{\ltap}{\>\centeron{\raise.35ex\hbox{$<$}}
                               {\lower.65ex\hbox{$\sim$}}\>}
\newcommand{\gtap}{\>\centeron{\raise.35ex\hbox{$>$}}
                               {\lower.65ex\hbox{$\sim$}}\>}

\newcommand\ZZ{\hbox{\zfont Z\kern-.4emZ}}
\font\zfont = cmss10 %scaled \magstep1
\newcommand{\sfrac}[2]{{\textstyle\frac{#1}{#2}}}

\begin{document}
\begin{titlepage}
\begin{flushright}
%{\tt hep-ph/yymmnn}
\end{flushright}

\vskip.5cm
\begin{center}
{\huge \bf 
Dimensions of Supersymmetric Operators from AdS/CFT
}

\vskip.1cm
\end{center}
\vskip0.2cm

\begin{center}
{\bf
{Giacomo Cacciapaglia}, {Guido Marandella},\\
{\rm and}
{John Terning}}
\end{center}
\vskip 8pt

\begin{center}
{\it
Department of Physics, University of California, Davis, CA  
95616.} \\
\vspace*{0.3cm}
{\tt  cacciapa@physics.ucdavis.edu,  maran@physics.ucdavis.edu, terning@physics.ucdavis.edu}
\end{center}

\vglue 0.3truecm

\begin{abstract}
\vskip 3pt
\noindent
We examine the AdS/CFT correspondence through a manifestly 5D supersymmetric formalism, corresponding to a 4D ${\mathcal N}=1$ supersymmetric CFT.
We find that the dimensions of scalar and fermionic component operators  are simply 
related, and that there is a smooth transition of scalar operator dimensions through the value $d_s = 2$.
By using this formalism, we also show that the formula used in the string literature for the dimension of fermion operators is incomplete.
\end{abstract}

\end{titlepage}

\newpage

%\renewcommand{\thefootnote}{(\arabic{footnote})}

%%%%%%%%%%%%%%%%%%%%%%%%%%%%%%%%%%%%%%%%%%%%%%%%%%%%%%
%%%%%%%%%%%%%%%%%%%%%%%%%%%%%%%%%%%%%%%%%%%%%%%%%%%%%%
\section{Introduction}
\label{sec:intro}
\setcounter{equation}{0}
\setcounter{footnote}{0}
%%%%%%%%%%%%%%%%%%%%%%%%%%%%%%%%%%%%%%%%%%%%%%%%%%%%%%
%%%%%%%%%%%%%%%%%%%%%%%%%%%%%%%%%%%%%%%%%%%%%%%%%%%%%%

The anti-de Sitter/conformal field theory (AdS/CFT) correspondence arose out of an understanding of 
${\mathcal N}=4$ supersymmetric (SUSY) Yang-Mills
theories using D-brane constructions in string theory \cite{maldacena}.  It has since been extended to 
CFTs with less supersymmetry and it is widely thought to extend even to non-SUSY CFTs. Most studies of
the correspondence have been performed using a non-SUSY formalism, making use of on-shell
component fields. However, for ${\mathcal N}=1$ SUSY CFTs the situation can be improved using a formalism for discussing SUSY 5D theories (which correspond to 4D ${\mathcal N}=2$ SUSY) in terms of 4D ${\mathcal N}=1$ superfields \cite{nima}.  Thus it should be straightforward
to study the AdS/CFT correspondence for  ${\mathcal N}=1$ SUSY CFTs keeping SUSY manifest 
throughout using off-shell auxiliary fields.  
We find that, by employing this formalism, we are able to
easily handle some subtle points~\cite{Witten} concerning scalar operators with dimensions below 2, which is
important for studying cases where the CFT operators are close to being free fields.
Moreover, keeping supersymmetry explicit helps to clarify the relation between scalar and fermion operator dimensions~\cite{contino}, and to show that formulas for the fermionic dimensions previously calculated in the literature~\cite{string} are incomplete.

The paper is structured as follows. 
In section 2 we review supersymmetric AdS$_5$ with hypermultiplets and the corresponding holographic boundary actions.
In Section 3 we discuss the CFT interpretation of these theories and extract the
dimensions of CFT operators corresponding to the bulk hypermultiplet.  
In section 4 we briefly describe how to extend the analysis to vector 
multiplets, and finally summarize our conclusions.

%%%%%%%%%%%%%%%%%%%%%%%%%%%%%%%%%%%%%%%%%%%%%%%%%%%%%%
%%%%%%%%%%%%%%%%%%%%%%%%%%%%%%%%%%%%%%%%%%%%%%%%%%%%%%
\section{SUSY in AdS space: chiral hypermultiplets}
\label{sec:susy}
\setcounter{equation}{0}
\setcounter{footnote}{0}
%%%%%%%%%%%%%%%%%%%%%%%%%%%%%%%%%%%%%%%%%%%%%%%%%%%%%%
%%%%%%%%%%%%%%%%%%%%%%%%%%%%%%%%%%%%%%%%%%%%%%%%%%%%%%

We start reviewing how to write a supersymmetic action in AdS$_5$ space.
We will work with the conformally flat metric:
\beq
d s^2 = \left( \frac{R}{z} \right)^2 \left( \eta_{\mu \nu} d x^\mu d x^\nu - d z^2 \right)\,,
\eeq
and define the theory on an interval delimited by $z_{UV}$ and $z_{IR}$.
We will recover the conformal limit by sending $z_{IR} \to \infty$ and $z_{UV} \to 0$. A 5D hypermultiplet consists of two 4D chiral superfields $\Phi$ and $\Phi_c$. The bulk action can be written in 4D superspace as~\cite{MartiPomarol}:
\beq
S & =& \int d^4 x\, d z \left\{ \int d^4 \theta\, \left( \frac{R}{z} \right)^3 \left[ \Phi^\ast\, \Phi + \Phi_c\, \Phi_c^\ast\right] + \right.\nonumber\\
& & \left.+ \int d^2 \theta\, \left( \frac{R}{z} \right)^3  \left[ \frac{1}{2}\; \Phi_c\, \partial_z \Phi - \frac{1}{2}\; \partial_z \Phi_c\, \Phi + \frac{c}{z}\; \Phi_c\, \Phi \right] + h.c. \right\}\,,
\eeq
which is explicitly hermitian without boundary terms.
Expanding in components, $\Phi=\{\phi, \chi, F\}$ and $\Phi_c =\{ \phi_c, \psi, F_c \}$, the action for the scalar (and auxiliary) components is:
\beq
S_{scalar} &=& \int d^4 x\, d z\;  \left( \frac{R}{z} \right)^3 \left\{ \partial_\mu \phi^\ast\, \partial^\mu \phi  + \partial_\mu \phi_c^\ast\, \partial^\mu \phi_c + F^\ast F + F_c^\ast F + \phantom{\frac{1}{2}}\right. \\
 & & \left. + \left[ \frac{1}{2}\; F_c\, \partial_z \phi - \frac{1}{2}\; \partial_z F_c\, \phi + \frac{c}{z}\; F_c\, \phi  + \frac{1}{2}\; \phi_c\, \partial_z F - \frac{1}{2}\; \partial_z \phi_c\, F + \frac{c}{z}\; \phi_c\, F + h.c. \right] \right\}\,. \nonumber 
\eeq

The scalars and $F-$components are coupled by derivatives along the extra dimension: therefore, we need to solve the two coupled equations of motion (EOMs) with appropriate boundary conditions, given by the minimization of the action, in the usual way\footnote{The results in Ref.~\cite{PomarolGherghetta} can be obtained if we assume an orbifold compactification and we integrate out the auxiliary fields~\cite{MartiPomarol}. Here we want to be more general, and define the theory on an interval. In our approach, the boundary terms in Ref.~\cite{PomarolGherghetta,MartiPomarol} are replaced by suitable boundary conditions.}.
Varying the action with respect to $F_c$ we get:
\beq
 \left( \frac{R}{z} \right)^3 \left[ F_c^\ast +  \partial_z \phi - \left( \frac{3}{2} - c\right) \frac{1}{z}\; \phi \right] \delta F_c\ - \frac{1}{2}\, \left( \frac{R}{z} \right)^3\, \left[ \delta F_c\; \phi \phantom{\frac{1}{2}} \hspace{-0.2cm} \right]^{z=z_{IR}}_{z=z_{UV}}  \eeq
where the last term is a boundary contribution which arises through an integration by parts.
The bulk EOM is:
\beq \label{eq:eomfc}
F_c^\ast = -  \partial_z \phi + \left( \frac{3}{2}-c \right) \frac{1}{z}\; \phi~.
\eeq
We can now calculate the second EOM by varying the action with respect to $\phi^*$:
\beq
\partial_\mu \partial^\mu \phi + \partial_z F_c^\ast - \left( \frac{3}{2} + c \right) \frac{1}{z}\; F_c^\ast =0\,.
\eeq
Using the EOM for $F_c$, Eq.\eq{eomfc}, we get
\beq \nonumber\\ 
\partial_\mu \partial^\mu \phi - \partial_z^2 \phi + \frac{3}{z}\; \partial_z \phi + \left( c^2 + c - \frac{15}{4} \right) \frac{1}{z^2}\; \phi   =  0\,.
\eeq
The scalar field $\phi$ therefore has an effective bulk mass
\beq
m^2 R^2 =c^2 + c - 15/4 .
\eeq
The solutions of this EOM are Bessel functions of order
\beq
\nu_L \equiv  \sqrt{4+m^2 R^2} = \left| c+\frac{1}{2} \right|\,.
\eeq

The boundary conditions (BCs) are determined by setting to zero the boundary contributions to the variation of the action.
The variation with respect of the two fields $F_c$ and $\phi$ will generate the following two terms:
\beq
\left. \frac{1}{2}\, \left( \frac{R}{z} \right)^3 \left[ F_c\; \delta \phi -  \delta F_c\; \phi\right]\right|_{z=z_{UV}}^{z=z_{IR}}\,.
\label{boundaryphi}
\eeq
Thus we see that both boundary variations vanish with either of the two possible boundary conditions:
\beq \label{eq:BC}
\phi = 0 \quad \mbox{or} \quad F_c = 0 = - \partial_5 \phi + \left(\frac{3}{2} - c \right) \frac{1}{z}\; \phi\,.
\eeq
The second equation, $F_c=0$, is equivalent to a BC for a localized scalar mass term and it is the same as the mass term for $\phi$ found in~\cite{PomarolGherghetta}.
If we want the BCs to be supersymmetric, Eqs.\eq{BC} correspond to either of the two chiral multiplets vanishing on the boundaries, $\Phi = 0$ or $\Phi_c = 0$.
Note also  that the effective boundary mass for $\phi$ in Eq.\eq{BC} is related to the bulk mass by
\beq
\frac{3}{2} - c = \left\{ \begin{array}{l}
2-\nu_L\; \mbox{for}\; c > -\frac{1}{2} \\
2+\nu_L\; \mbox{for}\; c<-\frac{1}{2}
\end{array} \right.\,.
\eeq

We can repeat the same exercise  for $\phi_c$ and $F$, finding:
\beq
F =  \partial_z \phi_c^\ast - \left( \frac{3}{2}+c \right) \frac{1}{z}\; \phi_c^\ast\,, 
\eeq
and an effective bulk mass for $\phi_c$
\beq
m_c^2 R^2 = c^2 - c - \frac{15}{4}~,  
\eeq
which determines the order of the Bessel functions in the solutions:
\beq
\nu_R \equiv  \sqrt{4+m_c^2 R^2} = \left| c-\frac{1}{2} \right|\,.
\eeq
The effective localized mass for $\phi_c$ is
\beq
\frac{3}{2} + c = \left\{ \begin{array}{l}
2-\nu_R\; \mbox{for}\; c < \frac{1}{2} \\
2+\nu_R\; \mbox{for}\; c>\frac{1}{2}
\end{array} \right.\,.
\eeq
The two chiral superfields are therefore related by taking $c \to -c$.

Regarding the fermionic sector, the situation is more straightforward, as they do not mix with any auxiliary fields.
The fermionic sector consists therefore of a 5D bulk fermion with bulk mass $m_f R= c$, thus the same analysis in~\cite{PomarolGherghetta,MartiPomarol} can be done.  The action can be written 
in terms of components as
\begin{equation} 
S = \int d^5 x 
\left(\frac{R}{z}\right)^4 
 \left( 
- i \bar{\chi}  \bar{\sigma}^\mu \partial_\mu \chi 
- i \psi  \sigma^{\mu} \partial_\mu \bar{\psi} 
+ \sfrac{1}{2} ( \psi \overleftrightarrow{\partial_z} \chi 
-  \bar{\chi}  \overleftrightarrow{\partial_z} \bar{\psi} )
+ \frac{c}{z} \left( \psi \chi + \bar{\chi} \bar{\psi} \right) 
\right), 
\end{equation} 
where $\overleftrightarrow{\partial_z}  = \overrightarrow{\partial_z}-\overleftarrow{\partial_z}$
with the convention that the differential operators act only on the spinors and not on the metric factors.
Note that the fermionic component $\chi$ and $\psi$ have been rescaled by a factor $\sqrt{R/z}$ to obtain the usual normalization~\cite{MartiPomarol}. We will perform the same rescaling when writing a chiral multiplet in components; however it is a matter of conventions and it does not play any important role in our discussion.

%%%%%%%%%%%%%%%%%%%%%%%%%%%%%%%%%%%%%%%%%%%%%%%%%%%%%%%%%%%%%
\subsection{Explicit solutions}
%%%%%%%%%%%%%%%%%%%%%%%%%%%%%%%%%%%%%%%%%%%%%%%%%%%%%%%%%%%%%

The bulk wave function solutions for fermions and scalars are:
\beq
\chi (p, z) & = & \chi_4(p)\; z^{5/2}\, \left( a \; J_{\nu_L} (p z) + b \; Y_{\nu_L} (p z) \right) \equiv \chi_4(p) \; z^{5/2}\, f_L (p z)\,, \\
\phi (p, z) & = & \phi_4(p)\; z^2\, f_L (p z)\,, \\
\psi (p, z) & = & \psi_4(p)\; z^{5/2}\, \left( a \; J_{\nu_R} (p z) + b \; Y_{\nu_R} (p z) \right) = \psi_4(p)\; z^{5/2}\, f_R (p z)\,, \\
\phi_c (p, z) & = & \phi_{c4}(p)\; z^2\, f_R (p z)\;
\eeq
where  $p = \sqrt{-\partial_\mu \partial^\mu}$, and the subscript ``$4$'' indicates 4 dimensional fields.
Note that the two 4D fermionic components are related by 4D Dirac equations:
\beq
-i \bar{\sigma}^\mu \partial_\mu \chi_4 + p \bar{\psi}_4= 0~,
\ \ \mathrm{and} \ \ 
-i \sigma^\mu \partial_\mu \bar{\psi}_4 + p\chi_4 = 0~. 
\eeq
Supersymmetry requires that the coefficients $a$ and $b$ are the same in all the solutions, while their ratio is fixed by the BCs on the IR brane (or asymptotic behavior at large $z$ in the $z_{IR} \to \infty$).
As an example, one can apply the BCs described above, and verify that they indeed enforce supersymmetry.
Let us check that imposing the BCs on the UV brane, independently of $a$ and $b$, leads to a supersymmetric spectrum.
The first choice is:
\beq
\Phi (z_{UV}) = 0 \Rightarrow \left\{ \begin{array}{l}
\chi (m, z_{UV} ) = 0\,, \\
\phi (m, z_{UV} ) = 0\,,\\
F (m, z_{UV} )  \propto  \left(\partial_z - \frac{( 3/2 + c )}{z_{UV}} \right) \phi_c (m, z_{UV} ) = 0\,.
\end{array} \right.
\eeq
 
The first two obviously imply $f_L (m, z_{UV}) = 0$. 
Regarding the third, it gives rise to the same spectrum once we observe that, due to some properties of the Bessel functions:
\beq
\left( \partial_z - \left( \frac{3}{2} + c \right) \frac{1}{z} \right) z^2 f_R (p, z) = p z^2 f_L (p, z)\,.
\eeq

For the other choice, $\Phi_c (z_{UV})=0$, we observe that, analogously
\beq
\left( \partial_z - \left( \frac{3}{2} - c \right) \frac{1}{z} \right) z^2 f_L (p, z) = - p z^2 f_R (p, z)\,,
\eeq
so that, in this case, the spectrum is given by $f_R (m, z_{UV})= 0$.

%%%%%%%%%%%%%%%%%%%%%%%%%%%%%%%%%%%%%%%%%%%%%%%%%%%%%%%%%%%%%
\subsection{Holographic Lagrangian}
%%%%%%%%%%%%%%%%%%%%%%%%%%%%%%%%%%%%%%%%%%%%%%%%%%%%%%%%%%%%%

In this section we want to compute the holographic action generated by the bulk solutions sketched in the previous section~\cite{perezvictoria,contino}.
As usual, we need to fix the value of one of the two superfields on the UV brane, i.e. either $\Phi(z_{UV}) = \Phi_0$ or $\Phi_c (z_{UV})= \Phi_0$, where $\Phi_0$ will play the role of the 4D superfield source in the holographic interpretation.
The first choice, $\Phi = \Phi_0$, can be achieved by adding the UV boundary superpotential term
\beq \label{eq:SUV}
S_{UV} = -\int d^4 x \,\frac{1}{2} \left( \frac{R}{z_{UV}} \right)^3 \left( \int d^2 \theta\,  \Phi_c (z_{UV})\, \Phi_0 + h.c. \right)\,.
\eeq
The variation of the scalar action on the UV brane is:
\beq
        \label{eq:scalarUV} 
\delta S_{scalar~UV}&=&- \frac{1}{2}  \left( \frac{R}{z_{UV}} \right)^3  \int d^4x 
 \left[ F_c \; \delta \phi + \phi_c\; \delta F  + \right. \nonumber \\
 &&  \left.\delta \phi_c\; (F_0 -F)  + \delta F_c (\phi_0-  \phi \right)]_{z = z_{UV}}+h.c.,
\eeq
while the variation of the fermion action on the UV brane is given by \cite{Csaki}:
\begin{equation} 
        \label{eq:fermUV} 
\delta S_{ferm~UV}= - \frac{1}{2}  \left( \frac{R}{z_{UV}} \right)^4  \int d^4x 
\left[  
\psi\; \delta \chi 
+ \delta \psi\; (\chi_0 -\chi)
\, \right]_{z = z_{UV}}+h.c. 
\end{equation} 

Requiring the variation of action on the UV boundary to vanish thus gives the BCs:
\beq \label{eq:BCholo}
\chi (z_{UV}) = \chi_0\,, \quad \phi(z_{UV}) = \phi_0\,, \quad F(z_{UV}) = F_0\,.
\eeq

Plugging the solutions of the EOMs back into the action, the bulk action vanishes due to the fact that all the EOMs are first order differential equations in $\partial_z$. Therefore the UV boundary term is~(\ref{eq:SUV})
\beq
S_{holo} = S_{UV} =-\frac{1}{2}\, \left( \frac{R}{z_{UV}}\right)^3 \int d^4 x\; \;  \left[  \frac{R}{z_{UV}}\; \psi\; \chi_0 + F_c\; \phi_0 + \phi_c\; F_0  \right]_{z=z_{UV}} + h.c.
\eeq

For the fermion fields, the normalizations of the bulk wave functions are fixed by the BC in Eq.~(\ref{eq:BCholo})
\beq
\chi (p,z)= \left( \frac{z}{z_{UV}} \right)^{5/2} \frac{f_L (p\, z)}{f_L (p\, z_{UV})}\, \chi_0(p)\,, \quad \bar{\psi} (p,z)= \left( \frac{z}{z_{UV}} \right)^{5/2} \frac{f_R (p\, z)}{f_L (p\, z_{UV})}\,\frac{p_\mu \bar{\sigma}^\mu}{p}\; \chi_0(p)\,. 
\eeq
Therefore, the fermionic holographic action is:
\beq
S_{holo} [\chi_0] = - \int d^4 x\; \left( \frac{R}{z_{UV}} \right)^4\; \bar{\chi}_0\;  \frac{f_R (p\, z_{UV})}{f_L (p\, z_{UV})}\,\frac{p_\mu \bar{\sigma}^\mu}{p}\; \chi_0\,. 
\eeq

For the scalar $\phi$:
\beq
\phi (p,z)=  \left( \frac{z}{z_{UV}} \right)^2 \frac{f_L (p\, z)}{f_L (p\, z_{UV})}\; \phi_0(p)
\eeq
and, using the EOM for $F_c$ in Eq.\eq{eomfc},
\beq
S_{holo} [\phi_0] &=& \int d^4 x \;\frac{1}{2} \left( \frac{R}{z_{UV}} \right)^3  \left[ \phi_0^\ast\, \left( \partial_z - \left( \frac{3}{2} - c \right) \frac{1}{z} \right) \phi + h.c. \right]_{z=z_{UV}} \nonumber \\
&=& - \int d^4 x \;
\left( \frac{R}{z_{UV}} \right)^3 \; \phi^\ast_0\; p\, \frac{f_R (p\, z_{UV})}{f_L (p\, z_{UV})}\; \phi_0\,.
\eeq

For the scalar $\phi_c$, from the BC
\beq
F (z_{UV}) = F_0 = \left. \left( \partial_z -\left( \frac{3}{2} +c \right) \frac{1}{z_{UV}} \right) \phi_c \right|_{z=z_{UV}}
\eeq
it follows that
\beq
\phi_c =  \frac{1}{p} \left( \frac{z}{z_{UV}} \right)^2 \frac{f_R (p\, z)}{f_L (p\, z_{UV})} F_0\,.
\eeq
The holographic action is therefore:
\beq
S_{holo} [F_0] = - \int d^4 x \; \left( \frac{R}{z_{UV}} \right)^3\; F^\ast_0\; \frac{1}{p}\, \frac{f_R (p\, z_{UV})}{f_L (p\, z_{UV})}\; F_0\,.
\eeq

We can now summarize the boundary action:
\beq \label{eq:holoac}
S_{holo} =  - \int d^4 x \;
\left[ \phi_0^\ast\; \Sigma_{\phi}\; \phi_0 + F_0^\ast\; \Sigma_{F}\; F + \chi_0^\ast\; \Sigma_{\chi}\; \chi_0 \right]\,;
\eeq
where the kinetic terms determined by
\beq \label{eq:holo}
\Sigma_{\phi} = \left( \frac{R}{z_{UV}} \right)^3\; p \frac{f_R}{f_L}\,, \qquad \Sigma_{\chi} = \left( \frac{R}{z_{UV}} \right)^4\; \frac{p_\mu \bar{\sigma}^\mu}{p} \frac{f_R}{f_L}\,, \qquad \Sigma_{F} = \left( \frac{R}{z_{UV}} \right)^3\; \frac{1}{p} \frac{f_R}{f_L}\,.
\eeq

In the case $\Phi_c (z_{UV}) = \Phi_0$, the UV boundary term is
\beq
S_{UV} =  \int d^4 x \frac{1}{2} \left( \frac{R}{z_{UV}} \right)^3 \left( \int d^2 \theta\, \Phi_0\; \Phi (z_{UV}) + h.c. \right)\,.
\eeq 
For the kinetic terms, up to an overall sign, we find the same expressions as in Eq.\eq{holo} with $L \leftrightarrow R$, i.e. $c \rightarrow -c$.

%%%%%%%%%%%%%%%%%%%%%%%%%%%%%%%%%%%%%%%%%%%%%%%%%%%%%%
%%%%%%%%%%%%%%%%%%%%%%%%%%%%%%%%%%%%%%%%%%%%%%%%%%%%%%
\section{CFT interpretation}
\label{sec:holo}
\setcounter{equation}{0}
\setcounter{footnote}{0}
%%%%%%%%%%%%%%%%%%%%%%%%%%%%%%%%%%%%%%%%%%%%%%%%%%%%%%
%%%%%%%%%%%%%%%%%%%%%%%%%%%%%%%%%%%%%%%%%%%%%%%%%%%%%%

We now want to give the CFT interpretation of the holographic action of Eq.\eq{holoac}. Since we have two scalar fields $\phi$ and $\phi_c$,  our naive intuition (based on non-SUSY results where there can be two different scaling dimensions for a given bulk field) would suggest that there are 4 possible scaling dimensions $d_s$ for the scalar sector of the CFT, which would be related to the scalar bulk masses by \cite{Witten}:
\beq \label{eq:witten} 
d_s = 
2 \pm \nu_L = \left\{ \begin{array}{l}
3/2 - c \\
5/2 + c
\end{array} \right. \quad \mbox{and}\quad d_s = 2 \pm \nu_R = \left\{ \begin{array}{l}
3/2 + c \\
5/2 - c
\end{array} \right.~.
\eeq
Depending on the value of $c$, some of those solutions will not be acceptable as they violate the unitarity bound $d_s > 1$.
However, as we will show in this section, not all of the remaining CFTs can be supersymmetric.

In our formalism, where the auxiliary F--components are included, it is important to correctly identify the propagators for the scalar and fermionic components of the CFT operator.
Consider  a chiral
superfield CFT operator $\Phi_\mathcal{O}$ with components $\{ {\mathcal O}, \Theta_{\mathcal O}, F_{\mathcal O} \}$.
The supersymmetric coupling between source and operator is a superpotential term of the form:

$$
\int d^2 \theta\; \Phi_\mathcal{O}\; \Phi_0 =  \Theta_\mathcal{O}\; \chi_0 + F_\mathcal{O}\; \phi_0 + \mathcal{O}\; F_0\,.
$$

The correct interpretation is that the source $F_0$ couples to the scalar component of the CFT, $\mathcal{O}$, so that the scalar correlator is
\beq 
\Delta_s (p) \equiv < \mathcal{O}(-p) \mathcal{O}(p) >  =\frac{\delta^2 S_{holo}}{\delta F_0(-p)\delta F_0(p)}= -\Sigma_F ( p )\,. 
\eeq
It is important to notice that it is the holographic action of the $F$--component, and not of the scalar, that contains information about the scalar CFT correlator.

On the other hand, the $F_\mathcal{O}$ component of the CFT supermultiplet couples with the scalar source $\phi_0$, therefore:
\beq \label{eq:2pfO}
\Delta_F (p) \equiv < F_\mathcal{O} (-p) F_\mathcal{O} (p) >  =\frac{\delta^2 S_{holo}}{\delta \phi_0(-p)\delta \phi_0(p)}= -\Sigma_\phi (p)\,.
\eeq

For the fermion $\Theta_\mathcal{O}$, as usual
\beq
 \Delta_f \equiv < \Theta_\mathcal{O} (-p) \Theta_\mathcal{O} (p) > =\frac{\delta^2 S_{holo}}{\delta \chi_0(-p)\delta \chi_0(p)}= - \Sigma_\chi (p)\,.
\eeq

The kinetic functions are given in Eq.\eq{holo}, and
are related to each other by:
\beq
\Delta_F = p^2 \Delta_s\,, \quad \Delta_f = p_\mu \bar{\sigma}^\mu\, \Delta_s\,;
\eeq
those relations are enough to ensure the correct relations between the scaling dimensions of the components of the supermultiplet, $d_s = d_f-1/2 = d_F - 1$, and this is a nice check that our interpretation is correct.

In order to extract the dimension of the chiral operator, we need to understand how the 2-point function in Eq.\eq{2pfO} scales with the momentum $p$.
It is a function of the product $p z_{UV}$: in the conformal limit $z_{UV} \to 0$, we can expand for small arguments $p z_{UV} \ll 1$.
Note that the 2-point function also depends on the ratio $a/b$, which is fixed by the asymptotic conditions at $z_{IR} \to \infty$ or by the BCs on the IR brane.
In the former case, $a/b$ is a number independent of the momentum.
In the case of a finite IR brane, $a/b$ does depend on $p$: however we are interested in the conformal limit $p z_{IR} \gg 1$ where the $a/b$ reduces to a ratio of trigonometric functions and it does not have any scaling with $p$.
To be more rigorous, we should extract the scaling properties of the residual at the poles corresponding to the KK tower of CFT bound states, however the conclusions would be the same.

Expanding the scalar propagator\footnote{Here we are assuming for simplicity that $c \pm 1/2$ is not an integer. In case of integers, logs arise from the expansion of the $Y$ Bessel functions.}, for small $p z_{UV}$:
\beq
\Delta_s & \sim & p^{-1} \frac{a (p\, z_{UV}) ^{|c-1/2|} + b (p\, z_{UV})^{-|c-1/2|}}{a (p\, z_{UV})^{|c+1/2|} + b(p\, z_{UV})^{-|c+1/2|}} + \dots \nonumber \\\
 & \sim & p^{|c+1/2| - |c-1/2| -1} \left( 1+\frac{a}{b} (p\, z_{UV})^{2|c+1/2|} -\frac{a}{b} (p\, z_{UV})^{2|c-1/2|} + \dots \right)
\eeq

For $c>1/2$, we get:
\beq
\Delta_s \sim \frac{1}{(z_{UV})^{2c-1}} ( 1 + \dots) +\frac{a}{b}\; (p)^{2c-1} + \dots
\eeq
where we have properly rescaled the correlator with powers of $z_{UV}$.
The dimension of the scalar operator is therefore $d_s = 3/2 + c = 2+\nu_R > 2$.
However, in the conformal limit $z_{UV} \to$~$0$, some of the local terms dominate.
Those terms can be canceled by adding a local supersymmetric action on the UV brane: this corresponds to the usual renormalization procedure.
Note also that the corresponding fermionic dimension, $d_f =d_s + 1/2 = 2+c$, agrees both with \cite{contino} and \cite{string}.

For $-1/2 < c < 1/2$:
\beq
\Delta_s \sim (p)^{2c-1} +\frac{a}{b}\; (z_{UV})^{1-2c} + \dots
\eeq
Now $d_s = 3/2+c = 2-\nu_R<2$, and the local terms vanish in the conformal limit. 
Note that in this range the dimension of the scalar is $1<d_s<2$: supersymmetry ensures a smooth transition between dimensions larger and smaller than 2. 
This is very different from the non-supersymmetric case where, in order to achieve $d_s<2$, it is necessary to change the BCs on the UV and Legendre--transform the action~\cite{Witten}: the modified BC is generated by a fine-tuned mass term on the UV-brane.  
The reason why a fine tuning is required for $d_s < 2$ is that a mass term (a bilinear operator in $\mathcal{O}$) becomes a relevant operator, and it has to be tuned away to keep the conformal symmetry unbroken for $d_s < 2$. 
In the supersymmetric case, scalar mass terms are protected by the chiral symmetry of their fermion superpartners, thus there is nothing special as $d_s$ goes below $2$. 
In the 5D model, the transition through $d_s=2$ is smooth as expected because supersymmetry takes care of generating the UV boundary condition which is imposed by hand in the non--supersymmetric case.
Note that in this region we still agree with \cite{contino}, but we disagree with the string formula for negative $c$.
As we will shortly see the formula in \cite{string} does describe the other choice of BCs for $c<0$.

Finally, for $c<-1/2$:
\beq
\Delta_s \sim  \frac{(1+\dots)}{p^{2}} + \frac{a}{b}\; (z_{UV})^{-2c-1} p^{-2c-3} + \dots
\eeq
Note that in this case the propagator has a pole: this pole cannot be canceled by a local term in the action on the UV brane, thus this signals the presence of an elementary field coupled to the source.
In the conformal limit, the non--local term vanishes.
We therefore interpret this case as a free field of canonical dimension $d_s = 1$; in other words the CFT operator is a free field that decouples (eg. this is what happens to the meson field operator in SUSY QCD for a sufficiently small number of flavors \cite{Seiberg}).
This interpretation is new compared to \cite{contino} and \cite{string}.
If we followed the analysis of the fermionic case in~\cite{contino},  we would interpret the non--local term as the contribution of an operator of dimension $d_s = 1/2-c$. 
However this interpretation relies on the presence of a finite UV brane and brane localized degrees of freedom which couple to the CFT\footnote{We thank R.~Contino for pointing this out to us.}.
In the limit $z_{UV} \to 0$ that we are considering, we are only left with a pure CFT.

For the other choice of BCs, it is enough to reverse the sign of $c$, therefore the result is
\beq
d_s = 3/2 - c\,, \quad d_f = 2 - c\, \quad \mbox{for} \quad c < 1/2\,,
\eeq
and free fields for $c>1/2$.
The formula $d_f =2+|c|$ used in the string literature \cite{string} only works when $c>0$  for the first choice of BCs, or $c<0$ for the second choice of BCs.  
The ``string" formula is clearly incomplete since it does not admit free fermions.

In summary, we have found that for every choice of bulk mass $c$, there are 2 CFTs depending on the BCs on the UV brane:
\beq
 \begin{array}{ccccl}
 \mbox{for} & c \ge1/2 & d_s = 3/2 + c\,,\; d_f = 2+c  & \mbox{or}&  d_s =1\,, \; d_f = 3/2  \\
\mbox{for} & -1/2\le c\le1/2 & d_s = 3/2 + c\,,\; d_f = 2+c  & \mbox{or}&  d_s =3/2-c\,, \; d_f = 2-c  \\
\mbox{for} & c\le-1/2 & d_s = 1\,,\; d_f = 3/2  & \mbox{or}&  d_s =3/2-c\,, \; d_f = 2-c  
\end{array} 
\eeq

In\cite{Witten}, the authors show how the AdS/CFT correspondence works for a scalar operator of dimension $1\le d_s < 2$: after modifying the UV boundary condition, a Legendre transformation is performed on
the holographic boundary action which exchanges the sources and the CFT operators (see also Ref. \cite{Mueck:1999jh}).
Under such a Legendre transformation, the kinetic operators in the holographic action are inverted. 
In the manifestly supersymmetric case we are analyzing, if we apply the Legendre transformation we would then identify the CFT correlators with the propagators of the sources in the following way 
\beq 
< \mathcal{O}(-p) \mathcal{O}(p) > = \Delta_s & = & (\Sigma_\phi)^{-1} \propto \frac{1}{p} \frac{f_L}{f_R}\,, \\
< F_\mathcal{O} (-p) F_\mathcal{O} (p) > = \Delta_F &=& (\Sigma_F )^{-1} \propto p  \frac{f_L}{f_R}\,, \\
< \Theta_\mathcal{O} (-p) \Theta_\mathcal{O} (p) > = \Delta_f & =& ( \Sigma_\chi )^{-1} \propto \frac{p^\mu \sigma_\mu}{p} \frac{f_L}{f_R}\,.
\eeq
These correlators correspond to the other choice of BCs $\Phi_c = \Phi_0$.
Thus the Legendre transformation simply interchanges the two choices of BCs.

%%%%%%%%%%%%%%%%%%%%%%%%%%%%%%%%%%%%%%%%%%%%%%%%%%%%%%
%%%%%%%%%%%%%%%%%%%%%%%%%%%%%%%%%%%%%%%%%%%%%%%%%%%%%%
\section{Vector supermultiplet}
\label{sec:vector}
\setcounter{equation}{0}
\setcounter{footnote}{0}
%%%%%%%%%%%%%%%%%%%%%%%%%%%%%%%%%%%%%%%%%%%%%%%%%%%%%%
%%%%%%%%%%%%%%%%%%%%%%%%%%%%%%%%%%%%%%%%%%%%%%%%%%%%%%

A vector multiplet in 5 dimensions can be described by a 4D vector multiplet $V = (A_\mu, \lambda_1, D)$ and a chiral multiplet $\chi = (\Sigma, \lambda_2, F)$.
The action can be written as~\cite{MartiPomarol}:
\beq
S & = & \int\, d^4 x\, dz\; \left\{ \frac{1}{4 g_5^2} \int d^2\theta\, W_\alpha W^\alpha + \frac{1}{g_5^2} \int d^4\theta\,  \left( \partial_z V - \frac{z}{R} \frac{\chi + \chi^\ast}{\sqrt{2}} \right)^2 \right\}\,.
\eeq
The EOMs for the auxiliary fields $F$ and $D$ yield:
\beq
F = 0\,, \qquad D = - \frac{R}{z} \left( \partial_z  - \frac{2}{z} \right) \Sigma\,,
\eeq
while the scalar and fermionic EOMs contain bulk masses~\cite{PomarolGherghetta} $m_\Sigma^2 R^2 = -4$ and $m_\lambda R = c_\lambda = 1/2$. 

Depending on the two possible BCs on the UV brane for the fermions, $\lambda_1 = 0$ or $\lambda_2 = 0$, the fermionic operator will have dimension $3/2$ or $5/2$; while the scalar operator has dimension $2$.
The first choice for the fermions correspond to the supersymmetric BCs $V = 0$ and $\chi = \chi_0$, does not lead to a vector superfield, but rather a chiral superfield with scaling dimension $2$.
The other choice is $\chi = 0$ and $V = V_0$; in this case the CFT operators are a vector and a fermion with canonical dimensions, i.e. one 4D vector supermultiplet.

%%%%%%%%%%%%%%%%%%%%%%%%%%%%%%%%%%%%%%%%%%%%%%%%%%%%%%
%%%%%%%%%%%%%%%%%%%%%%%%%%%%%%%%%%%%%%%%%%%%%%%%%%%%%%
\section{Conclusions}
\label{sec:concl}
\setcounter{equation}{0}
\setcounter{footnote}{0}
%%%%%%%%%%%%%%%%%%%%%%%%%%%%%%%%%%%%%%%%%%%%%%%%%%%%%%
%%%%%%%%%%%%%%%%%%%%%%%%%%%%%%%%%%%%%%%%%%%%%%%%%%%%%%

By maintaining manifest $\mathcal{N} = 1$ SUSY, we have seen that the usual calculations of operator dimensions in AdS/CFT are simplified.
Subtleties in the interpretation of scalar fields are avoided since scalar BCs automatically arise in a supersymmetric fashion through the auxiliary fields, rather than having to be imposed by hand. 
In this approach the behavior of scalar operators with dimensions below 2 arises naturally and $d_s = 2$ does not play any special role.
This is important because it clarifies how CFT operators transition toward free fields.

We also showed that for every value of the mass $c$ there are 2 different CFTs depending of the choice of boundary condition on the UV brane.
For a vector bulk field, the two choices are clearly independent: one is a vector multiplet of canonical dimension 1, the other choice lead to a chiral multiplet of dimension 2.
For a bulk hypermultiplet, the two BCs are related by a Legendre transformation.
Once the BC is fixed, varying the bulk mass $c$ the dimension of the operator decreases to the canonical dimension (for $c=1/2$ or $c=-1/2$); in the remaining range ($c>1/2$ or $c<-1/2$) the CFT operator reduces to a free field.

Moreover, supersymmetry relates the dimensions of the scalar and fermionic components of the CFT operator, therefore the calculations already present in the literature for non supersymmetric fields can be directly compared.
As a result, we showed that the formula for the fermion dimension $d_f = 2 + |c|$ used in the string literature is incomplete.

%%%%%%%%%%%%%%%%%%%%%%%%%%
\section*{Acknowledgments}
We thank Roberto Contino and Markus Luty for useful discussions and comments.
G.C. and J.T. also thank the Kavli Institute for Theoretical Physics in Santa Barbara, CA for hospitality during completion of this work.
This work is supported in part by the US Department of Energy under contract No. DE- 
FG03-91ER40674 and in part by the National Science Foundation under
Grant No. PHY05-51164.

\end{document}